\documentclass[11pt]{article}
\usepackage{jheppub}
\usepackage{axodraw}
\usepackage{bbm} 
\allowdisplaybreaks 

\newcommand{\la}[1]{\label{#1}}

\newcommand{\eq}{Eq.~}
\newcommand{\se}{Sec.~}
\newcommand{\app}{App.~}
\newcommand{\eqs}{Eqs.~}
\newcommand{\nr}[1]{(\ref{#1})}
\newcommand{\nn}{\nonumber \\}
\renewcommand{\(}{\left(}
\renewcommand{\)}{\right)}
\newcommand{\lb}{\left\{}
\newcommand{\rb}{\right\}}
\newcommand{\lk}{\left[}
\newcommand{\rk}{\right]}

\newcommand{\e}{\epsilon}
\newcommand{\order}[1]{{\cal O}(#1)}

\newcommand{\sumint}[1]{\hbox{$\sum$}\!\!\!\!\!\!\!\int_{#1}}
\newcommand{\sumintp}[1]{\hbox{$\sum^\prime$}\!\!\!\!\!\!\!\!\!\int_{#1}}
\newcommand{\sint}{\sum\!\!\!\!\!\!\int\;}

\newcommand{\gammaE}{{\gamma_{\small\rm E}}}

\newcommand{\intV}{{\cal M}_{1,0}}
\newcommand{\calVA}{{\cal A}}
\newcommand{\calVB}{{\cal B}}
\newcommand{\calVC}{{\cal C}}
\newcommand{\alVA}{A}
\newcommand{\alVB}{B}
\newcommand{\alVC}{C}
\newcommand{\intB}{{\cal M}_{0,0}}
\newcommand{\calA}{{\cal D}}
\newcommand{\calB}{{\cal E}}
\newcommand{\calC}{{\cal F}}
\newcommand{\alA}{D}
\newcommand{\alB}{E}
\newcommand{\alC}{F}
\newcommand{\intMbb}{{\cal M}_{2,-2}}
\newcommand{\Isqed}{{\cal I}_{\rm sqed}}
\newcommand{\T}{{\bar\Pi}}
\newcommand{\cA}{{\cal A}} 
\newcommand{\cB}{{\cal B}} 
\newcommand{\tr}{{\rm tr\,}}
\newcommand{\tP}{{\tilde\Pi}}
\renewcommand{\vec}[1]{{\mathbf{#1}}}

\newcommand{\pic}[1]{\;\parbox[c]{30pt}{\begin{picture}(30,30)(0,0)
\SetWidth{1.0}\SetScale{1.0} #1 \end{picture}}\;}
\newcommand{\picb}[1]{\;\parbox[c]{45pt}{\begin{picture}(45,30)(0,0)
\SetWidth{1.0}\SetScale{1.0} #1 \end{picture}}\;}
\def\scfc{0.8}  
\def\phgt{24}   
\def\pwc{24}    
\def\pwcb{36} 
\newcommand{\PIC}[4]{\;\parbox[c]{#2 pt}{\begin{picture}(#2,#3)(0,0)
\SetWidth{1.0}\SetScale{#4} #1 \end{picture}}\;}
\renewcommand{\pic}[1]{\PIC{#1}{\pwc}{\phgt}{\scfc}}
\renewcommand{\picb}[1]{\PIC{#1}{\pwcb}{\phgt}{\scfc}}
\def\ToptVS(#1,#2,#3){\pic{#1(15,15)(15,0,180) #2(15,15)(15,180,360)%
 #3(30,15)(0,15)}}
\def\ToprVB(#1,#2,#3,#4){\picb{#1(30,15)(15,-120,120) #2(30,15)(15,120,240)%
 #3(15,15)(15,60,300) #4(15,15)(15,-60,60)}}
\def\ToprVV(#1,#2,#3,#4,#5){\!\!\picb{#2(26.25,15)(15,256,76)%
 #3(30,30)(15,30) #1(18.75,15)(15,104,284) #4(15,30)(22.5,0)%
 #5(30,30)(22.5,0)}\!\!}
\def\ToprVMij(#1,#2,#3,#4,#5,#6){\pic{#3(15,15)(15,-30,90)%
 #1(15,15)(15,90,210)%
 #2(15,15)(15,210,330) #5(2,7.5)(15,15) #6(15,15)(15,30) #4(28,7.5)(15,15)%
 \Text(12.5,2)[b]{{$\scriptstyle i$}}\Text(13.5,18)[l]{{$\scriptstyle j$}}}}
\def\TopoSB(#1,#2,#3){\picb{#1(0,15)(7.5,15) #2(22.5,15)(15,0,180)%
 #3(22.5,15)(15,180,360) #1(37.5,15)(45,15)}}
\def\Lsc(#1,#2)(#3,#4){\Line(#1,#2)(#3,#4)}
\def\Asc(#1,#2)(#3,#4,#5){\CArc(#1,#2)(#3,#4,#5)}

\title{A fresh look on three-loop sum-integrals}

\preprint{BI-TP 2012/29}

\author{York Schr\"oder}

\affiliation{Faculty of Physics, University of Bielefeld,
  33501 Bielefeld, Germany}

\emailAdd{yorks@physik.uni-bielefeld.de}

\abstract{In order to prepare the ground for evaluating classes of three-loop
sum-integrals that are presently needed for thermodynamic observables,
we take a fresh and systematic
look on the few known cases, and review their evaluation
in a unified way using coherent notation. 
We do this for three important cases of massless bosonic three-loop 
vacuum sum-integrals that have been frequently used in the literature,
and aim for a streamlined exposition as compared to the original
evaluations. In passing, we speculate on options
for generalization of the computational techniques that have 
been employed.}

\begin{document}
\maketitle
\flushbottom

%
\section{Introduction}
\la{se:intro}

Our knowledge about sum-integrals, needed for evaluating
phenomenologically relevant
equilibrium observables in thermal field theories
(some examples being \cite{Arnold:1994eb,Braaten:1995jr,Laine:2005ai}),
is by far not as developed as the knowledge about 
continuum integrals \cite{Laporta:2001dd} needed for standard high-energy
(but zero-temperature) phenomenology, as e.g.\ reviewed
recently in \cite{Nishimura:2012ee}.
One of the reasons seems to be the impenetrable
structure of the multiple infinite sums that are involved.

Even considering the simplest class of sum-integrals,
dimensionally regularized 
massless bosonic vacuum sum-integrals (which 
constitute zero-scale problems since their 
only dimensionful scale -- the temperature $T$ -- scales out trivially) 
that we will call {\em hot tadpoles} in the following,
only a very limited number of cases are known in practice (for a
review, see \cite{Schroder:2008ex}),
let alone the number of technical tools that have been
developed for handling such sum-integrals.

All such presently known hot tadpoles contain
one-loop two-point sub-integrals $\Pi(P)$, whose structure is heavily
exploited in the process of evaluation.
While the 1-loop tadpoles can be computed exactly, 
i.e.\ as functions of the dimension $d$,
all 2-loop sum-integrals
can be factorized into 1-loop ones e.g.\ by systematic use of
integration-by-parts (IBP) methods \cite{Chetyrkin:1981qh,Nishimura:2012ee}, 
the first non-trivial
cases occur at the 3-loop level.

It turns out that the few hot 3-loop tadpoles that have been
evaluated in the literature have been dealt with on
a case-by-case basis by hand, often in some painstaking process,
involving inspired tensor transformations, 
elaborate (UV- and IR-) subtractions, skillful integration
tricks, mixed momentum and coordinate space techniques,
numerical integration and the 
such \cite{Arnold:1994ps,Gynther:2007bw,Andersen:2008bz}.

At present, however, there are not only pressing open
three-loop questions \cite{Moeller:2012da}, 
but even interesting open problems at four-loop order \cite{Kajantie:2000iz}
that involve a (large) number of yet unknown sum-integrals.
While exactly one type of 4-loop tadpole has been evaluated 
so far \cite{Gynther:2007bw},
it is clear that a more systematic treatment is urgently needed.
The purpose of the present note is to re-analyze the 
known non-trivial 3-loop cases and to streamline their derivations
in terms of a unified notation \cite{Moller:2010xw}, 
in order to prepare the ground for
tackling further 3-loop and 4-loop tadpoles in an efficient way.

The relevance of hot tadpoles can be appreciated from the following:
In modern treatments of equilibrium thermodynamics,
despite the known problem of infrared (IR) divergences in
massless thermal gauge theories \cite{Linde:1980ts},
an effective field theory (EFT) approach \cite{Ginsparg:1980ef}
allows for clean separation of IR and ultraviolet scales;
the sum-integrals treated here constitute, in the jargon of those
EFT's, the {\em hard} contributions to the corresponding observables, 
and hence need no IR regulator (other than being dimensionally
regularized) \cite{Braaten:1995cm}.

The three key examples of known non-trivial massless 3-loop 
vacuum sum-integrals 
that exhibit a range of useful techniques
are of `spectacles-', `basketball-' 
and `tensor-spectacles-type',
and can be represented as special instances of a general class 
as defined by 
(cf.\ \eqs(A.23),(A.30) of Ref.~\cite{Braaten:1995jr}), 
\begin{align}
\la{eq:Mclass}
\ToprVMij(\Asc,\Asc,\Asc,\Lsc,\Lsc,\Lsc) \equiv 
{\cal M}_{i,j} &\equiv \sumint{PQR} \frac{\lk(Q-R)^2\rk^{-j}}{\lk P^2\rk^i}\, 
\frac1{Q^2\,(Q+P)^2}\,\frac1{R^2\,(R+P)^2}
\;.
\end{align}
Our Euclidean notation is such that we use bosonic four-momenta $P$
with $P^2 = p_0^2 + \vec{p}^{2} = (2\pi n_p T)^2+\vec{p}^{2}$,
and where the sum-integral symbol
is a shorthand for
\begin{align}
\sumint{P} \equiv
T\sum_{p_{0}}\int\frac{\mathrm{d}^{d}\vec p}{(2\pi)^{d}}\,,
\end{align}
with $d=3-2\e$ 
and the sum taken over all integers $n_p\in\mathbbm{Z}$.
In the notation of \eq\nr{eq:Mclass}, $\intB$ and $\intV$ 
are the basketball- and spectacles-type 3-loop tadpoles, 
respectively, while we refer to $\intMbb$ as the tensor-spectacles-case,
owing to its numerator structure.

In the remainder of the paper, we will discuss the cases
$\intV$, $\intB$ and $\intMbb$ of \eq\nr{eq:Mclass} in turn, 
in Sections \ref{se:V}--\ref{se:M22}. \se\ref{se:conclu}
contains conclusions, while a few technical details are relegated to the 
appendices.

Before turning to the specific cases, let us note that
(by splitting $4QR=(Q+Q)(R+R)$ and exploiting the denominator's 
invariance by shifting $Q\rightarrow -Q-P$ or $R\rightarrow -R-P$
in the second instance only), the sub-class
\begin{align}
{\cal M}_{N,-1} =& 
\sumint{PQR} \frac{Q^2+R^2-4QR/2}{[P^2]^N\;Q^2\,(Q+P)^2\;R^2\,(R+P)^2}
\la{eq:MN1}
= 2I_1\sumint{P}\frac{\Pi(P)}{[P^2]^N}
-\frac12\sumint{P}\frac{[\Pi(P)]^2}{[P^2]^{N-1}}
\\&\mbox{with~~}
\Pi(P) \equiv \sumint{Q} \frac1{Q^2\,(P+Q)^2} \;=\; \TopoSB(\Lsc,\Asc,\Asc)
\end{align}
is seen to involve scalar 1-loop sub-integrals $\Pi$ only\footnote{To 
unclutter the text, we have collected various definitions in the 
Appendix (cf.\ \eq\nr{eq:Idef} for $I_1$).}. It can hence
be treated with the scalar methods employed for $\intV$ and $\intB$
see \se\ref{se:V} and \se\ref{se:B}, respectively.
As an example, from \eq\nr{eq:MN1} we immediately get
${\cal M}_{1,-1}=-\intB/2$, where we have used that the 2-loop sunset 
sum-integral
\begin{align}
\ToptVS(\Asc,\Asc,\Lsc)\equiv
{\cal S}\equiv\sumint{PQ}\frac{1}{P^2\,Q^2\,(Q+P)^2} \;=\;
\sumint{P}\frac{1}{P^2}\,\Pi(P) \;=\;0
\end{align}
vanishes identically in dimensional regularization as
can be shown via integration-by-parts (IBP) 
techniques \cite{Schroder:2008ex,Nishimura:2012ee}.

%
\section{The 3-loop spectacles}
\la{se:V}

In the notation of \eq\nr{eq:Mclass},
the 3-loop spectacles-type sum-integral $\intV$ is defined 
in terms of the 1-loop 2-point function $\Pi$ as 
\begin{align}
\la{eq:specDef}
\ToprVV(\Asc,\Asc,\Lsc,\Lsc,\Lsc)\equiv
\intV &\equiv \sumint{P} \frac1{P^2}\lk\Pi(P)\rk^2 \;.
\end{align}
We shall translate the original computation of $\intV$ 
from Ref.~\cite{Andersen:2008bz} (relying on the methods 
pioneered by \cite{Arnold:1994ps})
to our notation and systematics,
using less than their seven pages, in a transparent way,
and in a notation that is generalizable to other cases.

%
\subsection{Decomposition of $\intV$}

The spectacles can be identically re-written as
\begin{align}
\la{eq:Vrewritten}
\intV &= 
\sumint{P} \frac1{P^2}\lb 2\Pi_D\,\Pi\vphantom{\lk\Pi_A\rk^2}\rb
+\sumint{P} \frac{\delta_{p_0}}{P^2}\lb
\lk\Pi-\Pi_A\rk^2
+2\lk\Pi_A-\Pi_D\rk\Pi
-\lk\Pi_A\rk^2\rb
+\nn&+\sumintp{P} \frac1{P^2} \lb
\lk\Pi-\Pi_B\rk^2
+2\lk\Pi_B-\Pi_D\rk\lk\Pi-\Pi_C\rk
-\lk\Pi_B\rk^2 +2\lk\Pi_B-\Pi_D\rk\Pi_C \rb \;,
\end{align}
where $\delta_{p_0}$ picks out the Matsubara zero-mode, 
the primed sum excludes the zero-mode
and we have suppressed the argument $(P)$ of all functions 
in curly brackets.
This re-written form becomes useful if $\Pi_{A,B,C}$ have the
structure $\sum_i f_i(T,\e)\,/\,(P^2)^{n_i(\e)}$,
and $\Pi_D=f(T,\e)$ does not depend on the momentum $P$.
For then, the first term of \eq\nr{eq:Vrewritten} 
is proportional to the 2-loop sunset
sum-integral ${\cal S}$, which vanishes as already discussed above; 
all terms that do
not involve $\Pi$ are trivial 1-loop tadpoles $I$, which are known
analytically in $d=3-2\e$ dimensions (see \eq\nr{eq:Idef}); 
the zero-mode term involving $\lk\Pi_A-\Pi_D\rk\Pi$ 
is known analytically in terms of the 2-loop function $A$
introduced in \eq(21) of \cite{Moller:2010xw} (see \eq\nr{eq:Atad0});
and in the three remaining terms that involve
$\Pi$, the three subtraction terms $\Pi_{A,B,C}$ can be independently
chosen such as to facilitate their evaluation.

The strategy of Ref.~\cite{Andersen:2008bz} amounts to choosing\footnote{We take a
slightly different $\Pi_D$ here, whose leading term at
$\e\rightarrow0$ equals the choice $1/(16\pi^2\e)$ of \cite{Andersen:2008bz}.} 
(for a motivation of this choice, see below)
\begin{align}
\Pi_A = \frac{\beta}{(P^2)^\e}+\frac{T\,G(1,1,d)}{(P^2)^{\frac12+\e}}\;,\quad
\Pi_B = \frac{\beta}{(P^2)^\e}\;,\quad
\Pi_C = \frac{\beta}{(P^2)^\e}+\frac{2\,I_1}{P^2}\;,\quad
\Pi_D = \frac{\beta}{(\alpha\,T^2)^\e}\;,
\end{align}
where $\beta\equiv G(1,1,d+1)$ 
with $G$ given in \eq\nr{eq:Gdef}
and $\alpha$ is a constant to be fixed later.
\eq\nr{eq:Vrewritten} then reduces to
\begin{align}
\la{eq:Vdecomposed}
\intV &= 
0 
+\calVA 
+2\,\beta\,A(\overline{1\!+\!\e},1,1) 
+2\,T\,G(1,1,d)\,A(3/2+\e,1,1)
-0
\,+\nn&+\calVB +2\,\calVC -\beta^2\,I_{1+2\e} 
+2\,\beta^2\,I_{\overline{1\!+\!\e}}
+4\,\beta\,I_1\,I_{\bar{2}}
\;,
\end{align}
where we have introduced the shorthand
$f(\bar x)=f(x+\e)-f(x)/(\alpha\,T^2)^\e$
for convenience.
The first zero in \eq\nr{eq:Vdecomposed} is the 2-loop sunset discussed above and
the second zero comes from the fact 
that $\sint \delta_{p_0}\,\lk\Pi_A\rk^2/P^2$ 
is scale-free and hence vanishes in dimensional regularization.
We have introduced the notation $\calVA,\calVB,\calVC$ for the three non-trivial 
3-loop sum-integrals\footnote{In the notation of Ref.~\cite{Andersen:2008bz}, 
$\calVA=I_4^{b1}$,
$\calVB=I_4^a$ and 
$\calVC|_{\alpha=16\pi^2/e^\gammaE}= 
I_3^a|_{\mu=4\pi T/e^{(1+\gammaE/2)}}$.} 
\begin{align}
\calVA \equiv \sumint{P} \frac{\delta_{p_0}}{P^2} \lk\Pi-\Pi_A\rk^2\;,\quad
\calVB \equiv \sumintp{P} \frac1{P^2}
\lk\Pi-\Pi_B\rk^2\;,\quad
\calVC \equiv \sumintp{P} \frac1{P^2}
\lk\Pi_B-\Pi_D\rk\lk\Pi-\Pi_C\rk
\end{align}
that involve $\Pi$, and whose evaluation we shall discuss in the
next subsection.
In fact, the subtraction terms $\Pi_{A,B,C,D}$ defined above were 
chosen such that $\calVA,\calVB,\calVC$ are finite in $d=3$, and can be 
evaluated numerically after simplification by e.g.\ the spatial 
Fourier transform method of \cite{Arnold:1994ps}. 
More concretely,
$\Pi_A$ subtracts the leading UV- and IR-divergences in $\calVA$;
$\Pi_B$ subtracts the leading UV-divergence in $\calVB$;
$\Pi_D$ was chosen such that $\lk\Pi_B-\Pi_D\rk$ is
finite as $d\rightarrow3$; 
and $\Pi_C$ subtracts the leading and sub-leading UV-divergences 
in $\calVC$.

%
\subsection{Evaluation of $\calVA,\calVB,\calVC$}

Let us now bring the 3-loop sum-integrals $\calVA,\calVB,\calVC$ into
a form suitable for numerical evaluation. 
The (inverse) 3d spatial Fourier transforms that will be used below 
read \cite{Gynther:2007bw}
\begin{align}
\lb\Pi-\Pi_B,\frac{2\,I_1}{P^2}\rb &= 
\frac{T}{(4\pi)^2}\int\frac{{\rm d}^3\vec{r}}{r^2}\,
e^{i\vec{p}\vec{r}}\,e^{-|p_0|r}\lb\coth\(2\pi T r\)-\frac1{2\pi T r},
\frac{2\pi T r}{3}\rb +\order{\e}\;.
\end{align}

For $\calVA$, re-writing 
$\delta_{p_0}\lk\Pi-\Pi_A\rk=\delta_{p_0}\lk\Pi-\Pi_B\rk-\frac18\frac{T}{p}
\times\int\frac{{\rm d}^3\vec{r}}{r^2}\,
e^{i\vec{p}\vec{r}}\frac{8p}{(4\pi)^2} +\order{\e}$
(where $\frac18=G(1,1,3)$, while the extra integral is unity and introduced 
here for notational simplicity);
using the 3d spatial Fourier transform of $\lk\Pi-\Pi_B\rk$ (at $p_0=0$);
integrating over angles via 
$\frac12\int_{-1}^1{\rm d}u\,e^{i p r u}=\frac{\sin(pr)}{pr}$;
and letting $|\vec{r}|=x/(2\pi T)$, $|\vec{r}'|=y/(2\pi T)$, 
$|\vec{p}|=2\pi T z$:
\begin{align}
\calVA &= \frac{T^2}{(4\pi)^4}\,\alVA +\order{\e}\;,\\
\alVA &= 
\int_0^\infty \frac{{\rm d}x}{x}\(\coth(x)-\frac1x-1\)
\int_0^\infty \frac{{\rm d}y}{y}\(\coth(y)-\frac1y-1\)
\frac4\pi 
\int_0^\infty \frac{{\rm d}z}{z^2}\sin(zx)\sin(zy)
\nn
&= 
\int_0^\infty \frac{{\rm d}x}{x}\(\coth(x)-\frac1x-1\)
\int_0^\infty \frac{{\rm d}y}{y}\(\coth(y)-\frac1y-1\)
\(|x+y|-|x-y|\)
\nn
&= 2\int_0^\infty \frac{{\rm d}x}{x}\(\coth(x)-\frac1x-1\)
\int_0^x \frac{{\rm d}y}{y}\(\coth(y)-\frac1y-1\)
\(2y\)
\nn
&= 4\int_0^\infty \frac{{\rm d}x}x 
\(\coth(x)-\frac1x-1\)
\lk\ln\(\frac{\sinh(x)}x\)-x\rk
\nn
&= 2\int_0^\infty \frac{{\rm d}x}{x^2}
\lk\ln\(\frac{\sinh(x)}x\)-x\rk^2
\;\approx\; 9.5763057898\dots\;.
\end{align}

For $\calVB$, using the Fourier transform of $\lk\Pi-\Pi_B\rk$; 
recognizing $\int_{\vec{p}} e^{i\vec{p}\vec{r}}\frac1{\vec{p}^2+p_0^2}=
\frac{e^{-|p_0|r}}{4\pi r}$; 
summing over $p_0$ via geometric series;
scaling $|\vec{r}|=x/(2\pi T)$, $|\vec{r}'|=y/(2\pi T)$; 
and integrating over angles:
\begin{align}
\calVB &= \frac{T^2}{(4\pi)^4}\,\alVB +\order{\e}\;,\\
\alVB &= 4\int_0^\infty \frac{{\rm d}x}x \(\coth(x)-\frac1x\)
\int_0^x \frac{{\rm d}y}y \(\coth(y)-\frac1y\)
\lk\ln\(\frac{\sinh(x+y)}{\sinh(x)}\)-y\rk\\
&\approx 0.058739245719\dots\;.
\end{align}

For $\calVC$, using the Fourier transform of $\lk\Pi-\Pi_C\rk$;
expanding 
$\lk\Pi_B-\Pi_D\rk=\frac1{(4\pi)^2}\ln\frac{\alpha\,T^2}{P^2}+\order{\e}$;
integrating over angles; letting $|\vec{p}|=|p_0|y$, 
$|\vec{r}|=x/(2\pi T)$;
and using the exponential-integral
${\rm Ei}(x)\equiv -\int_{-x}^\infty \frac{{\rm d}t}t \,e^{-t}$
for 
$\frac2\pi\,e^{|z|}\int_0^\infty{\rm d}y\,\frac{y\,\sin(y|z|)}{y^2+1}\,
\ln\frac{\alpha}{y^2+1}=
e^{2|z|}{\rm Ei}(-2|z|)+\gammaE+\ln\frac{|z|\alpha}{2}$:
\begin{align}
\calVC &= -\frac13\,\frac{T^2}{(4\pi)^4}\,\alVC +\order{\e}\;,\\
\alVC &= -6\int_0^\infty \frac{{\rm d}x}x 
\(\coth(x)-\frac1x-\frac{x}3\)
\sum_{n=1}^\infty \lk {\rm Ei}(-2nx)+e^{-2nx}\,
\ln\(\frac{2x}n\frac{\alpha e^\gammaE}{16\pi^2}\)\rk
\\
&\approx 0.003496\dots\;.
\end{align}
where the numerical value is given for $\alpha=16\pi^2/e^\gammaE$
and corresponds to\footnote{Note that for this choice of $\alpha$, we
avoid the computation of $\xi$ in \eqs(D.20-25) in \cite{Andersen:2008bz}.} 
\eq(D.27) of \cite{Andersen:2008bz}. For an discussion of
the numerical evaluation, we refer to \app\ref{se:numVABC}.

%
\subsection{Result}

Expanding \eq\nr{eq:Vdecomposed} around $d=3-2\e$ 
(for $\alpha=16\pi^2/e^\gammaE$), 
we finally obtain 
\begin{align}
\intV &= -\frac14\, \frac{T^2}{(4\pi)^4}\, 
\frac{\(4\pi e^\gammaE T^2\)^{-3\e}}{\e^2} 
\lk 1 +v_1\,\e + v_2\,\e^2 +\order{\e^3}\rk\;,\\
v_1 &= \frac43+4\gammaE+2\frac{\zeta'(-1)}{\zeta(-1)}\;,\\
\la{eq:v2}
v_2 &= \frac13\lk 46-16\gammaE^2+\frac{45\pi^2}4+24\ln^2(2\pi)
-104\gamma_1-8\gammaE-24\gammaE\ln(2\pi)
+\right.\nn&\left.{}
+16\gammaE\frac{\zeta'(-1)}{\zeta(-1)} 
+24\frac{\zeta'(-1)}{\zeta(-1)} 
+2\frac{\zeta''(-1)}{\zeta(-1)} \rk
-38.5309\dots\;,
\end{align}
which coincides with \eq(D.51) of \cite{Andersen:2008bz}.
The numerical value in \eq\nr{eq:v2} is $-4\( \alVA+\alVB-\frac23\,\alVC\)$.

%
\section{The 3-loop basketball}
\la{se:B}
 
In the notation of \eq\nr{eq:Mclass},
the basketball-type sum-integral $\intB$ is defined 
in terms of the 1-loop 2-point function $\Pi$ as 
\begin{align}
\ToprVB(\Asc,\Asc,\Asc,\Asc)\equiv
\intB &\equiv \sumint{P} \lk\Pi(P)\rk^2 \;.
\end{align}
Historically, the evaluation of $\intB$ was performed by 
in Ref.~\cite{Arnold:1994ps}, where
many of the techniques that were later generalized to 
other cases of sum-integrals, such as basketball-type 
tadpoles with different powers on the 
propagators and/or factors in the 
numerator \cite{Moller:2010xw,Moeller:2012da}, were introduced.
Here, we translate this pioneering computation of $\intB$ 
to our notation and systematics.

%
\subsection{Decomposition of $\intB$}

The basketball can be identically re-written as
\begin{align}
\la{eq:Brewritten}
\intB &= 
\sumint{P} \lb 2\,\Pi_D\,\Pi\vphantom{\lk\Pi_A\rk^2}\rb
+\sumint{P} \delta_{p_0}\lb
\lk\Pi\!-\!\Pi_B\rk^2
+2\lk\Pi_B\!-\!\Pi_D\rk\Pi
-\lk\Pi_B\rk^2\rb
+\sumintp{P} \lb 2\lk\Pi_C\!-\!\Pi_B\rk\Pi\vphantom{\lk\Pi_A\rk^2}\rb
+\nn&
+\sumintp{P} \lb
\lk\Pi-\Pi_C\rk^2
+2\lk\Pi_B-\Pi_D\rk\lk\Pi-\Pi_E\rk 
-\lk\Pi_C\rk^2 +2\lk\Pi_B-\Pi_D\rk\Pi_E \rb \;,
\end{align}
where, in full analogy to \eq\nr{eq:Vrewritten},
$\delta_{p_0}$ picks out the Matsubara zero-mode, 
the primed sum excludes the zero-mode
and we have suppressed the argument $(P)$ of all functions 
in curly brackets.

Let us slightly refine the strategy of Ref.~\cite{Arnold:1994ps} by choosing
\begin{align}
\la{eq:below}
\Pi_B = \frac{\beta}{(P^2)^\e}\;,\;\;
\Pi_C = \Pi_B\!+\!\frac{2\,I_1}{P^2}\;,\;\;
\Pi_D = \frac{\beta}{(\alpha\,T^2)^\e}\;,\;\;
\Pi_E = \Pi_C\!+\!8\,\frac{T^4 J_1}{[P^2]^2}\,\frac{P^2\!-\!(d\!+\!1)p_0^2}{d\,P^2}
\;,
\end{align}
where $\beta\equiv G(1,1,d+1)$ as above, 
$\alpha$ is a constant to be fixed later
and $J_n=\(\frac{4\pi}{T^2}\)^\e
\frac{\Gamma(d+n)\zeta(d+n)}{4\pi^{3/2}\Gamma(d/2)}$ as in 
\eq(B5) of \cite{Arnold:1994ps}.
With this choice, 
the first term of \eq\nr{eq:Brewritten} can be shifted to the square 
of a trivial 1-loop tadpole $I$;
for the last term in the first line of \eq\nr{eq:Brewritten}, 
note that after $\sum^\prime\rightarrow\sum-\sum\delta_{p_0}$
the full sum is proportional to the 2-loop sunset
sum-integral ${\cal S}$ and hence vanishes; 
the rest as well as the other zero modes involving $\lk\Pi_B-\Pi_D\rk\Pi$ and 
$\lk\Pi_B-\Pi_C\rk\Pi$ 
are known analytically in terms of the 2-loop function $A$
of \eq\nr{eq:Atad0};
all terms that do
not involve $\Pi$ are trivial 1-loop tadpoles $I$; 
and in the three remaining terms that involve
$\Pi$, the subtraction terms $\Pi_{B,C,D,E}$ have been
chosen such as to subtract UV divergences 
in order to render the sum-integrals finite.

\eq\nr{eq:Brewritten} then reduces to
\begin{align}
\la{eq:Bdecomposed}
\intB &= 
2\,\beta\,I_1\,I_1/(\alpha\,T^2)^{\e}
+\calA 
+2\,\beta\,A(\bar 0,1,1) 
-0+0-4\,I_1\,A(1,1,1)
\,+\\\nonumber&+\calB +2\,\calC 
-\beta^2\,I_{2\e} -4\,\beta\,I_1\,I_{1+\e}-4\,I_1\,I_1\,I_2
+2\,\beta^2\,I_{\bar{\e}}
+4\,\beta\,I_1\,I_{\bar 1}
+\frac{16\,\beta\,J_1}{d\,T^{-4}}\lk I_{\bar 2}\!-\!(d\!+\!1)I_{\bar 3}^2\rk
\;,
\end{align}
where we have again used the shorthand
$f(\bar x)=f(x+\e)-f(x)/(\alpha\,T^2)^\e$.
The first zero in \eq\nr{eq:Bdecomposed} comes from the fact 
that $\sum\!\!\!\!\!\!\int\; \delta_{p_0}\,\lk\Pi_B\rk^2$ 
is scale-free and hence vanishes in dimensional regularization
and the second zero is the 2-loop sunset discussed above.
We have introduced the notation $\calA,\calB,\calC$ for the three non-trivial 
3-loop sum-integrals\footnote{In the notation of Ref.~\cite{Arnold:1994ps}, 
$\calA=\mbox{\eq}\!(2.34)$;
$\calB\sim\mbox{\eq}\!(2.31,32)$; 
$\calC\sim I_a$ as treated in \eqs\!(D9-D14) of \cite{Arnold:1994ps}.} 
\begin{align}
\la{eq:calABC}
\calA \equiv \sumint{P} \delta_{p_0}\lk\Pi-\Pi_B\rk^2\;,\quad
\calB \equiv \sumintp{P}
\lk\Pi-\Pi_C\rk^2\;,\quad
\calC \equiv \sumintp{P}
\lk\Pi_B-\Pi_D\rk\lk\Pi-\Pi_E\rk
\end{align}
that involve $\Pi$, and whose evaluation we shall discuss in the
next subsection.

%
\subsection{Evaluation of $\calA,\calB,\calC$}

The 3-loop sum-integrals $\calA,\calB,\calC$ are finite as $d\rightarrow3$, 
such that the 3d spatial Fourier transform method of \cite{Arnold:1994ps}
proves fruitful.
The (inverse) transforms needed below are \cite{Arnold:1994ps,Gynther:2007bw}
\begin{align}
&\lb\Pi,\;\Pi_B,\;\Pi_C-\Pi_B,\;\Pi_E-\Pi_C,\;\Pi_B-\Pi_D\rb = \nn
&=\frac{T}{(4\pi)^2}\int\frac{{\rm d}^3\vec{r}}{r^2}\,
e^{i\vec{p}\vec{r}}\,e^{-|p_0|r}
\lb \coth\(\bar{r}\)+\frac{|p_0|}{2\pi T},\;
\frac{|p_0|}{2\pi T}+\frac1{\bar{r}},\;
\frac{\bar{r}}{3},\;
-\frac{\bar{r}^3}{45},\;
\frac{1+|p_0|r}{\bar{r}}\rb 
+\order{\e}\;,
\end{align}
where $\bar{r}=2\pi T r$, and for the last term we have expanded 
$\lk\Pi_B-\Pi_D\rk=\frac1{(4\pi)^2}\ln\frac{\alpha\,T^2}{P^2}+\order{\e}$
and used the inverse Fourier transform of the logarithm as derived
e.g.\ in \eqs(D.11),(D.12) of \cite{Arnold:1994ps}\footnote{Note that it is defined up to a Delta function 
$\delta(\vec{r})$ which however vanishes in $\calC$ below.}.

For $\calA$, 
using the 3d spatial Fourier transform of $\lk\Pi-\Pi_B\rk$ (at $p_0=0$);
integrating over angles;
and letting $|\vec{r}|=x/(2\pi T)$:
\begin{align}
\calA &= \frac{T^4}{(4\pi)^2}\,\alA +\order{\e}\;,\\
\label{eq:A}
\alA &= 
\frac12\int_0^\infty \frac{{\rm d}x}{x^2}\(\coth(x)-\frac1x\)^2
\;=\; \frac{2\,\zeta(3)}{\pi^2} \;,
\end{align}
where the analytic value was obtained via the recursion of  
\app\ref{se:recu}.

For $\calB$, using the Fourier transform of $\lk\Pi-\Pi_C\rk$; 
integrating over angles;
summing over $p_0$ via geometric series
and re-writing $2/(e^{2x}-1)=\coth(x)-1$;
and scaling $|\vec{r}|=x/(2\pi T)$:
\begin{align}
\calB &= \frac{T^4}{(4\pi)^2}\,\alB +\order{\e}\;,\\
\label{eq:B}
\alB &= \frac12\int_0^\infty \frac{{\rm d}x}{x^2} \(\coth(x)-\frac1x-\frac{x}3\)^2
\(\coth(x)-1\)\\
&= \frac1{18}\lk\frac{67}{30}+\gammaE-6\ln(2\pi)
-\frac{36\zeta(3)}{\pi^2}
-2\frac{\zeta'(-3)}{\zeta(-3)}
+7\frac{\zeta'(-1)}{\zeta(-1)}
\rk\;,
\end{align}
where the analytic value was again obtained via the recursion of 
\app\ref{se:recu}.

For $\calC$, 
using the Fourier transforms of $\lk\Pi-\Pi_E\rk$ and $\lk\Pi_B-\Pi_D\rk$;
integrating over angles;
letting $p_0=2\pi T n$, 
$|\vec{r}|=x/(2\pi T)$;
and summing over $n$ via geometric series
and re-writing $2/(e^{2x}-1)=\coth(x)-1$:
\begin{align}
\calC &= \frac{T^4}{(4\pi)^2}\,\alC +\order{\e}\;,\\
\label{eq:C}
\alC &= \frac12 \int_0^\infty \frac{{\rm d}x}{x^3} 
\(\coth(x)-\frac1x-\frac{x}3+\frac{x^3}{45}\)
\(1-\frac{x}2\,\partial_x\)\(\coth(x)-1\)
\\
\la{eq:calC}
&= \frac1{180}\lk-\frac{23}6+3\gammaE+\frac{90\zeta(3)}{\pi^2}
+2\frac{\zeta'(-3)}{\zeta(-3)}
-5\frac{\zeta'(-1)}{\zeta(-1)}
\rk\;,
\end{align}
where the analytic value was obtained by first introducing the regulator
$x^\delta$ as in \app\ref{se:recu}; 
re-writing $2\coth(x)\,\partial_x\coth(x)=\partial_x\coth^2(x)$;
integrating by parts all terms involving $\partial_x$ while dropping
all boundary terms, which vanish due to the regulator; 
and using the recursion of \app\ref{se:recu}, 
letting $\delta\rightarrow0$ in the end.

%
\subsection{Result}

Expanding \eq\nr{eq:Bdecomposed} around $d=3-2\e$ 
($\alpha$ does not contribute yet), 
we finally obtain 
\begin{align}
\intB &= \frac{T^4}{(4\pi)^2}\, 
\frac{\(4\pi e^\gammaE T^2\)^{-3\e}}{24\,\e} 
\lk 1 +b_{11}\,\e +\order{\e^2}\rk\;,\\
b_{11} &= \frac{37}9-\frac{32\gammaE}{15}+8\ln(2\pi)-\frac{24\zeta(3)}{\pi^2}
+\frac2{15}\frac{\zeta'(-3)}{\zeta(-3)}+24(\alA+\alB+2\alC)\nn
&= \frac{91}{15}
-2\frac{\zeta'(-3)}{\zeta(-3)}
+8\frac{\zeta'(-1)}{\zeta(-1)}
\;,
\end{align}
which coincides with \eq(2.36) of \cite{Arnold:1994ps}.

%
\section{The 3-loop tensor spectacles}
\la{se:M22}

A first non-trivial representative of the class \eq\nr{eq:Mclass}
involving numerator structure is $\intMbb$. 
In this case, we do not have an easy way of dealing with the 
scalar products in the numerator as was still the case for ${\cal M}_{1,-1}$, 
see \eq\nr{eq:MN1}.
Here, we wish to first relate the computation of $\intMbb$
to an auxiliary one \cite{Braaten:1995jr},
and then evaluate the latter, for historical reasons 
denoted $\Isqed$ \cite{Arnold:1994ps},
using our notation and systematics.
Let us note that $\intMbb$ is quite another 
category compared the previous two,
as it needs tensor methods.
Let us re-write\footnote{Note that in our conventions $d=3-2\e$ and hence
the trace of the metric tensor is $g_{\mu\mu}=d+1$.}
\begin{align}
\la{eq:M22dec1}
&4\intMbb = 4(5-d)I_2I_1I_1 +\intB
+\sumint{P} \frac1{[P^2]^2}\lk\T_{\mu\nu}(P)\rk^2\\
\la{eq:Pibar}
&{\rm with}\quad \T_{\mu\nu}(P) \equiv 2I_1 g_{\mu\nu}
-\sumint{Q}\frac{(2Q+P)_\mu(2Q+P)_\nu}{Q^2\,(Q+P)^2} \;,
\end{align}
where $\T_{\mu\nu}(P)$ was chosen transverse, $P_\mu\T_{\mu\nu}(P)=0$,
consequences of which will be exploited next.

%
\subsection{Relating $\intMbb$ to $\Isqed$}
\la{se:M2I}

Transversality $P_\mu \T_{\mu\nu}(P)=0$ 
constrains the structure of the symmetric tensor $\T_{\mu\nu}$ to
\begin{align}
\la{eq:PiAB}
\T_{\mu\nu}(P) &= {\cA}_{\mu\nu}\T_{\cA}(P)+{\cB}_{\mu\nu}\T_{\cB}(P) \;,
\end{align}
where $\cA_{\mu\nu}=\cA_{\nu\mu}$, $\cB_{\mu\nu}=\cB_{\nu\mu}$
are projectors $\cA\cA=\cA$, $\cB\cB=\cB$, $\cA\cB=0$
(with traces $\tr\cA=d-1$, $\tr\cB=1$)
which are orthogonal to the external momentum $P\cA=0=P\cB$.
Concretely, trading the 4-vector $U=(1,\vec 0)$ for the
linear combination $V\equiv P^2 U-(PU)P$
that satisfies $PV=0$, we have $\cA=g-\frac{PP}{P^2}-\frac{VV}{V^2}$
(for which in fact also $V\cA=0=U\cA$)
and $B=\frac{VV}{V^2}$.
The scalar coefficients $\T_{\cA,\cB}$ of \eq\nr{eq:PiAB}
can hence be obtained via projections\footnote{Clearly, 
$\T_{\cB}$ will be the ``hard'' case whenever it occurs,
involving $1/\vec{p}^2$ and $q_0$ etc.}
\begin{align}
\la{eq:projA}
\T_{\cA}(P) &= \frac{\tr\cA\T(P)}{\tr\cA\cA} 
= \frac1{d-1}\(g_{\mu\nu}-\frac{[P^2]^2}{V^2}U_\mu U_\nu\)\T_{\mu\nu}(P)
=\frac{\T_{\mu\mu}(P)-\T_{\cB}(P)}{d-1}\;,\\
\T_{\cB}(P) &= \frac{\tr\cB\T(P)}{\tr\cB\cB} 
= \frac{[P^2]^2}{V^2}\,U_\mu U_\nu\T_{\mu\nu}(P)
= \frac{P^2}{\vec{p}^2}\,\T_{00}(P) \;.
\end{align}

For the specific case at hand, 
we can read off $\T_{\mu\mu}$ and $\T_{00}$ from \eq\nr{eq:Pibar}
to obtain
\begin{align}
\la{eq:PiAPiB}
\T_{\mu\mu} = P^2\Pi(P)+2I_1(d-1) \;,\quad
\T_{\cB}(P) = \frac{P^2}{\vec{p}^2}
\bigg(2I_1-\sumint{Q}\frac{(2q_0+p_0)^2}{Q^2\,(Q+P)^2}\bigg) \;.
\end{align}

Returning to \eq\nr{eq:M22dec1}, 
$\intB$ is the 3-loop basketball sum-integral,
while the last term is related to $\Isqed$ 
of Appendix H in \cite{Arnold:1994ps} by an 
IR subtraction in the $p_0=0$ mode defined by
\begin{align}
\la{eq:PiIR}
\T_{\mu\nu}^{\rm IR} &\equiv  
{\cA}_{\mu\nu}\T_{\cA}^{\rm IR}+{\cB}_{\mu\nu}\T_{\cB}^{\rm IR}
\end{align}
when choosing\footnote{This particular choice in fact reflects 
$\T_{\{\cA,\cB\}}^{\rm IR} = \T_{\{\cA,\cB\}}(0,\vec{p}\rightarrow0)$.}
$\T_{\cA}^{\rm IR} = \frac{2(d-2)I_1+4I_2^2}{d-1}$ and
$\T_{\cB}^{\rm IR} = 2I_1-4I_2^2$ as momentum-independent:
\begin{align}
\sumint{P} \frac1{[P^2]^2}\lk\T_{\mu\nu}(P)\rk^2 &= 
\sumint{P} \frac1{[P^2]^2}
\lk\T_{\mu\nu}(P)-\delta_{p_0}\T_{\mu\nu}^{\rm IR}\rk^2 
+\sumint{P} \frac{\delta_{p_0}}{[P^2]^2}
\lk2\T_{\mu\nu}(P)-\T_{\mu\nu}^{\rm IR}\rk \T_{\mu\nu}^{\rm IR}
\nn
&= \Isqed 
+2\T_{\cA}^{\rm IR}\sumint{P} 
\frac{\delta_{p_0}}{[P^2]^2}\T_{\mu\mu} 
+2\(\T_{\cB}^{\rm IR}-\T_{\cA}^{\rm IR}\)\sumint{P} 
\frac{\delta_{p_0}}{[P^2]^2}\T_{\cB}
+0_{\rm scale-free}
\nn
\la{eq:M22dec2}
&= \Isqed 
+2\T_{\cA}^{\rm IR}A(1,1,1;0)
+8\(\T_{\cA}^{\rm IR}-\T_{\cB}^{\rm IR}\)A(2,1,1;2) 
+0
\;.
\end{align}
Here, the 1st line is a trivial re-writing;
the 2nd line follows via \eqs\nr{eq:PiAB},\nr{eq:PiIR}, 
using the properties of the projectors, plugging in \eq\nr{eq:projA},
noting that $\T_{\{\cA,\cB\}}^{\rm IR}$ are momentum-independent and dropping 
scale-free integrals that vanish in dimensional regularization;
and in the last line we have used \eq\nr{eq:PiAPiB}, 
again dropped scale-free integrals
and expressed the remaining sum-integrals
in terms of the 2-loop tadpole $A$ from \eq\nr{eq:Atad}.

%
\subsection{Evaluation of $\Isqed$}

Owing to \eqs\nr{eq:M22dec1} and \nr{eq:M22dec2}, 
instead of 
$\intMbb$ the authors of Ref.~\cite{Arnold:1994ps} choose to 
compute
\begin{align}
\Isqed &\equiv \sumint{P} \frac1{[P^2]^2}
\lk\T_{\mu\nu}(P)-\delta_{p_0}\T_{\mu\nu}^{\rm IR}\rk^2 \\
&= \sumint{P} \frac1{[P^2]^2} \lb
\lk\T_{\mu\nu}-\T_{\mu\nu}^{\rm UV}-\delta_{p_0}\T_{\mu\nu}^{\rm IR}\rk^2
+\T_{\mu\nu}^{\rm UV}\lk2\T_{\mu\nu}-\T_{\mu\nu}^{\rm UV}
-2\delta_{p_0}\T_{\mu\nu}^{\rm IR}\rk \rb \\
&= \sumint{P} \Bigg\{ \frac{\big[\tP_A-\tP_B\big]^2}{d-1}
+\lk\tP_B\rk^2 +\frac{\T^{\rm UV}}{d\,P^2}\Big(2\T_{\mu\mu}-P^2\T^{\rm UV}
-2\delta_{p_0}\lk(d-1)\T_{\cA}^{\rm IR}+\T_{\cB}^{\rm IR}\rk\Big) \Bigg\}
\nn
\la{eq:H14}
&= \frac1{d-1}\sumint{P} \lb 
\tP_B \lk d\,\tP_B-2\,\tP_A\rk
 +\lk\tP_A\rk^2\rb
+\frac1d\sumint{P} \T^{\rm UV}\bigg(\frac{2\T_{\mu\mu}}{P^2}-\T^{\rm UV}\bigg)
+0
\;,
\end{align}
where in the 2nd line a UV subtraction 
$\T_{\mu\nu}^{\rm UV}=
\({\cA}_{\mu\nu}+{\cB}_{\mu\nu}\)\frac{P^2}{d}\T^{\rm UV}$ 
was introduced\footnote{Note that its tensor structure is 
$({\cA}_{\mu\nu}+{\cB}_{\mu\nu})=(g_{\mu\nu}-P_\mu P_\nu/P^2)$,
as expected at zero temperature.}; for the 3rd line we have used
projector properties as well as \eq\nr{eq:projA}, and defined
\begin{align}
\tP_A &=\frac1{P^2}\(\T_{\mu\mu}-P^2\T^{\rm UV}
-\delta_{p_0}\lk(d-1)\T_{\cA}^{\rm IR}+\T_{\cB}^{\rm IR}\rk\) \;,\\
\tP_B &=\frac1{P^2}\(\T_{\cB}-P^2\T^{\rm UV}/d
-\delta_{p_0}\T_{\cB}^{\rm IR}\)\;;
\end{align}
and for the 4th line we have assumed $\T^{\rm UV}=\sum_i c_i/[P^2]^{n_i}$ 
and dropped scale-free integrals.

The essence of the computation of Ref.~\cite{Arnold:1994ps} 
is now the treatment of the terms 
involving $\tP_B$ in \eq\nr{eq:H14} 
(note the similarity to \eq(H.14) of \cite{Arnold:1994ps}).
Choosing different UV subtractions for the zero- and non-zero modes
via $\T^{\rm UV}=\Pi_B+(1-\delta_{p_0})d\,2I_1/P^2$
these terms are finite and can be treated in $d=3$ by 
spatial Fourier transform methods.
Ref.~\cite{Arnold:1994ps} states the simple result
\begin{align}
\la{eq:essence}
\frac1{d-2}\sumint{P}  
\tP_B \lk d\,\tP_B-2\,\tP_A\rk &=
\sumint{P}\lk\tP_A\rk^2+\order{\e} \;,
\end{align}
which follows
from a rather lengthy calculation, 
involves an ``amazing cancellation''
and is explained in our Appendix \ref{se:derivation}.
As a result,
the computation of $\Isqed$ (up to the constant term)
is reduced to elements that already appear in the basketball case $\intB$.

In detail, with \eq\nr{eq:PiAPiB} and the choices of $\T^{\rm IR}$ 
and $\T^{\rm UV}$ given above,
\begin{align}
\la{eq:PiA}
\tP_A &= \Pi-\Pi_B-(1-\delta_{p_0})\,2I_1/P^2
\;,\\
\la{eq:PiB}
\tP_B &= \frac1{P^2}\(\T_{\cB}-P^2\Pi_B/d-2I_1+\delta_{p_0}4I_2^2\)\;.
\end{align}
Plugging \eq\nr{eq:essence} into \eq\nr{eq:H14} then results in
\begin{align}
\Isqed &= \sumint{P}\lk\tP_A\rk^2
+\frac1d\sumint{P} \T^{\rm UV}\bigg(\frac{2\T_{\mu\mu}}{P^2}-\T^{\rm UV}\bigg)
+\order{\e}\\
\la{eq:IsqedYS}
&= \sumint{P}\Big\{ 
\!\lk\Pi\rk^2
+\frac{d\!-\!1}d \Big( \!\lk4I_1/P^2\!+\!\Pi_B\!-\!2\Pi\rk\Pi_B 
+(1\!-\!\delta_{p_0})d\!\lk2I_1/P^2\rk^2 \Big)
\Big\}\!+\!\order{\e}\;.
\end{align}
Recognizing in \eq\nr{eq:IsqedYS} the term quadratic in $\Pi$ as 
$\intB$, re-writing the linear term as
\begin{align}
\sumint{P}\Big\{\Pi\,\Pi_B\Big\} &= 
\sumintp{P}\Big\{
\lk\Pi-\Pi_E\rk\lk\Pi_B-\Pi_D\rk+\Pi_E\lk\Pi_B-\Pi_D\rk\Big\}
+\nn&
+\sumint{P}\delta_{p_0}\Big\{\Pi\lk\Pi_B-\Pi_D\rk\Big\}
+\sumint{P}\Big\{\Pi\,\Pi_D\Big\}
\end{align}
where the first term on the right-hand side (rhs) is $\calC$ (of the $\intB$ calculation
of \se\ref{se:B}, cf.\ \eqs\nr{eq:calABC},\nr{eq:calC}) 
and the others are elementary 
(as are the remaining terms in \eq\nr{eq:IsqedYS}),
$\Isqed$ evaluates to 
(using again $f(\bar x)=f(x+\e)-f(x)/(\alpha\,T^2)^\e$
as well as the functions collected in \eq\nr{eq:below} 
and in \app\ref{se:ints})
\begin{align}
\la{eq:Idecomposed}
\Isqed &= \intB 
+\frac{d\!-\!1}d\bigg(
\beta^2\,I_{2\e}
+4\,\beta\,I_1\,I_{1+e}
-2\Big\{
\calC
+\beta^2\,I_{\bar{\e}}
+2\,\beta\,I_1\,I_{\bar 1}
+\frac{8\,\beta\,J_1}{d\,T^{-4}}\lk I_{\bar 2}\!-\!(d\!+\!1)I_{\bar 3}^2\rk
{}+\nn&+
\beta\,A(\bar 0,1,1;0)
+\beta\,I_1\,I_1/(\alpha\,T^2)^\e
\Big\}
+4\,d\,I_2\,I_1\,I_1
+0_{\rm scale-free}
\bigg) +\order{\e}\;.
\end{align}
The expansion around $d=3-2\e$ coincides with \eq(H.30) 
of \cite{Arnold:1994ps}:
\begin{align*}
\Isqed &= \frac{T^4}{(4\pi)^2}\,
\frac{\(4\pi e^{\gammaE} T^2\)^{-3\e}}{108} 
\lk\frac{23}{2\,\e}+\lk\frac{517}{10}+12\gammaE
-11\frac{\zeta'(-3)}{\zeta(-3)}+68\frac{\zeta'(-1)}{\zeta(-1)}\rk
+\order{\e}\rk\;.
\end{align*}

%
\subsection{Result}

Putting together \eqs\nr{eq:M22dec1}, \nr{eq:M22dec2}, \nr{eq:Idecomposed}
and expanding around $d=3-2\e$, we finally obtain
\begin{align}
\intMbb &= \frac{T^4}{(4\pi)^2}\, 
\frac{\(4\pi e^\gammaE T^2\)^{-3\e}}{216}
\lk \frac{11}{\e}  +\lk\frac{73}{2}+12\gammaE
-10\frac{\zeta'(-3)}{\zeta(-3)}+64\frac{\zeta'(-1)}{\zeta(-1)}\rk 
+\order{\e}\rk\;,
\end{align}
which coincides with \eq(A.30) of \cite{Braaten:1995jr}.

%
\section{Conclusions}
\la{se:conclu}

We have re-examined the three most prominent cases of massless 
bosonic three-loop vacuum sum-integrals, in order to simplify
their derivation and translate the original calculations
to a language that is amenable to generalizations.

First, we have re-derived the result for the spectacles-type 3-loop vacuum
sum-integral given first by Andersen and Kyllingstad in 
\cite{Andersen:2008bz}, streamlining the computation quite a bit by 
using our notation from \cite{Moller:2010xw}. 
As an improvement over \cite{Andersen:2008bz}, 
we give a one-dimensional integral representation of $\alVA$ 
(which was given as a triple integral there).
Further effort would be welcome in order to derive a high-precision
result for the numerical coefficient $\alVC$, involving an infinite sum and
a one-dimensional integral, leading to extremely slow convergence behavior.
It would be interesting to study generalizations of the
computation outlined in \se\ref{se:V}, such as $1/P^2\rightarrow1/[P^2]^N$
as was done for the 3-loop basketball topology in \cite{Moller:2010xw},
or including factors of $p_0$ or other scalar products in the numerator,
in order to derive some of the integrals needed in our 3-loop computations.

Second, we have re-derived the result for the basic basketball-type 3-loop vacuum 
sum-integral given first by Arnold and Zhai in 
\cite{Arnold:1994ps}, streamlining the computation quite a bit by 
using our notation from \cite{Moller:2010xw}. 

Third, we have re-derived the result for the first non-trivial 3-loop vacuum
sum-integral involving scalar products in the numerator given first
by Arnold and Zhai in \cite{Arnold:1994ps}, 
somewhat streamlining the computation. 
Here is a summary of this computation in a nutshell: 
$$\intMbb \!\sim\! (QR)^2 \!=\! Q_\mu Q_\nu R_\mu R_\nu 
\stackrel{\rm rewrite}{\longrightarrow} [\Pi_{\mu\nu}]^2
\stackrel{\rm orthog}{\longrightarrow} [\Pi_{00},\Pi_{\mu\mu}]^2
\stackrel{\rm 3d\,FT}{\longrightarrow} [\Pi_{\mu\mu}]^2\!+\!\order{\e}
\!\sim\! \intB \!+\!\order{\e}$$
One wonders whether there is a simpler way 
to compute $\intMbb$.
Note that the projection method, acting on the level of sub-integrals,
seems to over-complicate the computation by triggering factors of
$1/\vec{p}^2$ (stemming from $1/V^2=1/P^2\vec{p}^2$), which
leads outside the class of integrals \eq\nr{eq:Mclass} we started with.
It even leads outside the natural generalization of this class 
as suggested by IBP methods (which allows for factors of $q_0$ etc 
in the numerators \cite{Nishimura:2012ee}). 
One idea to avoid this change of structure could be to explore
applicability of the generic tensor method of Ref.~\cite{Tarasov:1996br}
to the case of finite-temperature sum-integrals as discussed here. 
However, this is clearly beyond the scope of the present paper but should be 
explored in the future.

In closing, we hope that our unified exposure of known
techniques for sum-integral evaluation leads to a program
of generalizing them to other cases -- be it with irreducibles
in the numerator or with different powers of the 
denominators -- as needed for example for determining matching
coefficients in effective field theories \cite{Moeller:2012da}, or 
for advancing to the next loop order \cite{Gynther:2007bw}.
In the short term, it seems that the class of hot bosonic tadpoles
${\cal M}_{N,-2}$ is a suitable candidate deserving further study.
Finally, an extension to fermionic cases (ultimately involving masses
as well as chemical potentials) would be another possible line
of future work.

%
\section*{Acknowledgements}

Y.S.~is supported by the Heisenberg program of the Deutsche
Forschungsgemeinschaft (DFG), contract no.~SCHR~993/1. 
All diagrams were drawn with Axodraw~\cite{Vermaseren:1994je}.


\begin{appendix} 

%
\section{Standard integrals}
\la{se:ints}

For convenience, we collect here the functions used above, as defined
in \cite{Moller:2010xw}. 
They are the 1-loop massless propagator at zero temperature 
\begin{align}
\la{eq:Gdef}
G(s_1,s_2,d) &\equiv \(p^2\)^{s_{12}-\frac{d}2}\int\frac{{\rm d}^dq}{(2\pi)^d}
\frac1{[q^2]^{s_1}[(p+q)^2]^{s_2}}
=
\frac{\Gamma(\frac{d}2-s_1)\Gamma(\frac{d}2-s_2)\Gamma(s_{12}-\frac{d}2)}
{(4\pi)^{d/2}\Gamma(s_1)\Gamma(s_2)\Gamma(d-s_{12})} \;,
\end{align}
the 1-loop bosonic tadpoles 
\begin{align}
\la{eq:Idef}
I_s^a \equiv \sumint{Q} \frac{|q_0|^a}{[Q^2]^s} 
= \frac{2T\,\zeta(2s-a-d)}{(2\pi T)^{2s-a-d}}\,
\frac{\Gamma(s-\frac{d}2)}{(4\pi)^{d/2}\Gamma(s)}
\;,\quad
I_s \equiv \sumint{Q} \frac1{[Q^2]^s} 
= I_s^0\;,
\end{align}
and a specific 2-loop tadpole
\begin{align}
\la{eq:Atad0}
A(s_1,s_2,s_3) &\equiv A(s_1,s_2,s_3;0) \\
\la{eq:Atad}
A(s_1,s_2,s_3;s_4) &\equiv \sumint{PQ} 
\frac{\delta_{q_0}|q_0|^{s_4}}{[Q^2]^{s_1}[P^2]^{s_2}[(P+Q)^2]^{s_3}}
\nn&=
\frac{2T^2\,\zeta(2s_{123}-2d-s_4)}{(2\pi T)^{2s_{123}-2d-s_4}}\,
\frac{\Gamma(s_{13}-\frac{d}2)\Gamma(s_{12}-\frac{d}2)
\Gamma(\frac{d}2-s_1)\Gamma(s_{123}-d)}
{(4\pi)^d\Gamma(s_2)\Gamma(s_3)\Gamma(d/2)\Gamma(s_{1123}-d)}
\;,
\end{align}
where $s_{abc...} \equiv s_c+s_b+s_c+...\;$.

%
\section{Numerical evaluation of $\alVA,\alVB,\alVC$}
\la{se:numVABC}

The integrals $\alVA$ and $\alVB$ are easily evaluated 
numerically e.g.\ with Mathematica\cite{mma},
\begin{align}
\alVA &\approx 9.5763057898113125\;,\\
\alVB &\approx 0.058739245719225247\;,
\end{align}
while $\alVC$ is tougher to get with high precision. 
Maybe it is easier to handle it in pieces:
\begin{align}
\alVC_1 &= -6\int_1^\infty \frac{{\rm d}x}x 
\(\coth(x)-\frac1x-\frac{x}3\)
\sum_{n=1}^\infty \lk {\rm Ei}(-2nx)+e^{-2nx}\,\ln\(\frac{2x}n\)\rk\nn
&\approx +0.016232689597 \;,
\nn
\alVC_2 &= -6\int_0^1 \frac{{\rm d}x}x 
\(\coth(x)-\frac1x-\frac{x}3\)
\sum_{n=1}^\infty \lk {\rm Ei}(-2nx)-e^{-2nx}\,\ln\(2nx\,e^\gammaE\)\rk\nn
&\approx -0.022965150204 \;,
\nn
\alVC_3 &= -6\int_0^1 \frac{{\rm d}x}x 
\lk\(\coth(x)-\frac1x-\frac{x}3\)\frac1{e^{2x}-1}+\frac{x^2}{90}\rk
\ln\(4x^2 e^\gammaE\)\nn
&\approx -0.021888498587 \;,
\nn
\alVC_4 &= +6\int_0^1 \frac{{\rm d}x}x 
\frac{x^2}{90} \ln\(4x^2 e^\gammaE\)
\;=\; \frac{\gammaE+2\ln(2)-1}{30}\nn
&\approx +0.032117000867 \;,
\nn
\alVC &= \alVC_1+\alVC_2+\alVC_3+\alVC_4\nn
&\approx +0.003496041673 \;.
\end{align}
The sum in $\alVC_1$ converges reasonably fast, 
the one in $\alVC_2$ rather slowly.
The following piece of Mathematica code was used to obtain
the approximate numerical results given above:
\begin{verbatim}
wp = 60; ag = 30; max = 10000;

(*here the sum converges reasonably fast*)
Clear[cc1];
cc1[n_] := 
  cc1[n] = NIntegrate[-6/
      x (Coth[x] - 1/x - x/3) (ExpIntegralEi[-2 n x] + 
       Exp[-2 n x] Log[2 x/n]), {x, 1, Infinity}, 
    WorkingPrecision -> wp, AccuracyGoal -> ag];
c1 = Sum[cc1[n], {n, 1, max}]

(*this is the part in which the sum converges extremely slowly*)
Clear[cc2];
cc2[n_] := 
  cc2[n] = NIntegrate[-6/
      x (Coth[x] - 1/x - x/3) (ExpIntegralEi[-2 n x] - 
       Exp[-2 n x] Log[2 n x Exp[EulerGamma]]), {x, 0, 1}, 
    WorkingPrecision -> wp, AccuracyGoal -> ag];
c2 = Sum[cc2[n], {n, 1, max}]

(*these are not problematic at all*)
c3 = 
 NIntegrate[-6/
    x ((Coth[x] - 1/x - x/3)/(Exp[2 x] - 1) + x^2/90) (2 Log[2 x] + 
     EulerGamma), {x, 0, 1}, WorkingPrecision -> wp, 
  AccuracyGoal -> ag]
c4 = Integrate[x/15 (2 Log[2 x] + EulerGamma), {x, 0, 1}]

(*sum up to get C*)
{c1, c2, c3, N[c4, wp]}
Total[%]
\end{verbatim}

%
\section{Integrals needed for $\alA,\alB,\alC$}
\la{se:recu}

Following  Appendix C of Ref.~\cite{Arnold:1994ps},
let us introduce -- in a manner similar to dimensional regularization -- 
a convergence factor $x^\delta$ into convergent integrals of the
form \eqs\nr{eq:A}, \nr{eq:B} and \nr{eq:C}, enabling us to integrate 
term by term and letting $\delta\rightarrow0$ in the end.
The basic relations needed are
\begin{align}
H(a,b) &\equiv \int_0^\infty{\rm d}x\,x^{a+\delta}\,\coth^b(x) \;,\\
H(a,0) &= \int_0^\infty{\rm d}x\,x^{a+\delta} = 0 \;,\\
H(a,1) &= 
\int_0^\infty{\rm d}x\,x^{a+\delta} \lk 1+2\sum_{n=1}^\infty e^{-2nx} \rk
= H(a,0) +\frac{\Gamma(1+a+\delta)\,\zeta(1+a+\delta)}{2^{a+\delta}}\;,\\
H(a,b) &= \dots [ipb] \dots = \frac{a+\delta}{b-1}\,H(a-1,b-1)+H(a,b-2) \;,
\end{align}
allowing for a recursive solution of integrals 
$\int_0^\infty{\rm d}x\,x^z \coth^b(x)$ for $b\in\mathbb{Z}$.

%
\section{Derivation of \eq\nr{eq:essence}}
\la{se:derivation}

In order to derive \eq\nr{eq:essence}, we closely follow
\eqs(H.15-27) of Ref.~\cite{Arnold:1994ps}.
The (inverse) 3d spatial Fourier transforms that we need
derive from \eq(22) of \cite{Moller:2010xw}, 
\begin{align}
\frac{\Gamma(s)}{\lk P^2\rk^s} &= 
\frac{T}{(4\pi)^2}\!\int\!\frac{{\rm d}^3\vec{r}}{r^2}\,
e^{i\vec{p}\vec{r}}\,e^{-|p_0|r} 
\lb \frac{(4\pi)^2}{2^s\,(2\pi T)^{2s}}\,
\sum_{n=0}^{\hat s} \frac{(\hat s\!+\!n)!}{(\hat s\!-\!n)!} \,
\frac{|\bar{p}_0|^{1-s-n}\,\bar{r}^{s-n}}{2^n\,n!}
\rb +\order{\e}
\end{align}
where $\hat s=|3/2-s|-1/2$, and read
(using $\beta/\Gamma(\e)=1/(4\pi)^2+\order{\e}$
and $I_1=T^2/12+\order{\e}$)
\begin{align}
\la{eq:FTs}
&\lb\frac1{P^2},\frac{2I_1}{P^2},\Pi_B,P^2\Pi_B,
\int_{\vec q}\frac1{Q^2(Q+P)^2}\rb =
\frac{T}{(4\pi)^2}\int\frac{{\rm d}^3\vec{r}}{r^2}\,
e^{i\vec{p}\vec{r}}\,e^{-|p_0|r} 
\times\nn&\times
\lb\frac{2\bar r}{T^2},\frac{\bar r}{3},|\bar p_0|+\frac1{\bar r},
-2(2\pi T)^2\(\frac{\bar p_0^2}{\bar r}+\frac{3|\bar p_0|}{\bar r^2}
+\frac{3}{\bar r^3}\),e^{-(|q_0|+|q_0+p_0|-|p_0|)r}\rb 
+\order{\e}\;.
\end{align}
\mbox{}From \eqs\nr{eq:PiB} and \nr{eq:PiAPiB} we can therefore compute
the transform of \eq\nr{eq:PiB}:
\begin{align}
\vec{p}^2\,\tP_B &= 
-\sumint{Q}\frac{4q_0^2-p_0^2}{Q^2\,(Q+P)^2}
-\frac{P^2\,\Pi_B}{d}
+p_0^2\lk\frac{\Pi_B}{d}+\frac{2I_1}{P^2}\rk
+\delta_{p_0}\,4\,I_2^2
\nn
&= \frac{T}{(4\pi)^2}\int\frac{{\rm d}^3\vec{r}}{r^2}\,
e^{i\vec{p}\vec{r}}\,e^{-|p_0|r} 
(2\pi T)^2 \bigg\{
-\lk |\bar{p}_0|\frac{2+\bar{p}_0^2}3+\bar{p}_0^2\coth(\bar r)
+2\frac{\coth(\bar r)+|\bar{p}_0|}{\sinh^2(\bar r)}\rk
+\nn&+
\frac23\lk\frac{\bar{p}_0^2}{\bar r}+\frac{3|\bar{p}_0|}{\bar r^2}
+\frac{3}{\bar r^3}\rk
+\bar{p}_0^2\lk\frac13\(|\bar{p}_0|+\frac1{\bar r}\)+\frac{\bar r}3\rk
\bigg\}-\delta_{p_0}\frac{T^2}6+\order{\e} 
\nn
&=
-\frac{T}{(4\pi)^2}\int\frac{{\rm d}^3\vec{r}}{r^2}\,
e^{i\vec{p}\vec{r}}\,\partial_r^2\,e^{-|p_0|r}
\( \coth\(\bar{r}\)-\frac1{\bar{r}}-\frac{\bar{r}}{3}\)
-\delta_{p_0}\frac{T^2}6+\order{\e}
\nn
\la{eq:A6}
&=
-\frac{T}{4\pi}\int_0^\infty\!\!\!{\rm d}r\,
e^{-|p_0|r}
\( \coth\(\bar{r}\)-\frac1{\bar{r}}-(1-\delta_{p_0})\frac{\bar{r}}{3}\)
\partial_r^2\,\frac{\sin(pr)}{pr}
+\order{\e}\;,
\end{align}
where in the first line we have transformed the numerator of the 
sum-integral in \eq\nr{eq:PiAPiB} as 
$(2q_0+p_0)^2=4q_0^2+p_0^2+2(q_0+q_0)p_0\rightarrow4q_0^2-p_0^2$
(shifting half of the term linear in $q_0$ as $Q\rightarrow-Q-P$
using the denominator's symmetry) and re-arranged terms;
in the second line we have applied \eq\nr{eq:FTs}, 
solved the Matsubara 
sum\footnote{From \eq(19) of \cite{Moller:2010xw}, 
it immediately follows that $\sum_{q_0}e^{-(|q_0|+|q_0+p_0|-|p_0|)r}=
\lk\coth(\bar r)+|\bar p_0|\rk$ and\\that
$\sum_{q_0}2\bar q_0^2\,e^{-(|q_0|+|q_0+p_0|-|p_0|)r}=
\lk|\bar p_0|(1+2\bar p_0^2)/3
+\bar p_0^2\coth(\bar r)
+(\coth(\bar r)+|\bar p_0|)/\sinh^2(\bar r)\rk$.} 
and used $I_2^2=-T^2/24+\order{\e}$;
as a third step we have 
done an identical re-writing in terms of derivatives;
and in the last line we have 
integrated over angles via 
$\int\frac{{\rm d}^3\vec{r}}{r^2}\,e^{i\vec{p}\vec{r}}=
4\pi\int_0^\infty\!\!\!{\rm d}r\,\frac{\sin(pr)}{pr}$
and integrated by parts (twice; surface terms cancel 
against $\delta_{p_0}T^2/6$).
Analogously, for \eq\nr{eq:PiA} one easily gets
\begin{align}
\la{eq:A4}
\tP_A
&=
\frac{T}{(4\pi)^2}\int\frac{{\rm d}^3\vec{r}}{r^2}\,
e^{i\vec{p}\vec{r}}\,e^{-|p_0|r}
\( \coth\(\bar{r}\)-\frac1{\bar{r}}-(1-\delta_{p_0})\frac{\bar{r}}{3}\)
+\order{\e}\\
\la{eq:A5}
&=
\frac{T}{4\pi}\int_0^\infty\!\!\!{\rm d}r\,
e^{-|p_0|r}
\( \coth\(\bar{r}\)-\frac1{\bar{r}}-(1-\delta_{p_0})\frac{\bar{r}}{3}\)
\frac{\sin(pr)}{pr}
+\order{\e}\;.
\end{align}

To prove \eq\nr{eq:essence}, let us now use \eq\nr{eq:A4} on its 
rhs and integrate over angles:
\begin{align}
\sumint{P} \lk\tP_A\rk^2 
&= T\sum_{p_0} \frac{T^2}{(4\pi)^4}\int\!\frac{{\rm d}^3\vec{r}}{r^4}\,
e^{-2|p_0|r}\(\coth(\bar{r})-\frac1{\bar{r}}-(1-\delta_{p_0})\frac{\bar{r}}3\)^2
+\order{\e}\\
\la{eq:target}
&=  \frac{T}{4\pi} \sum_{p_0} \int_0^\infty\!\!\!{\rm d}r\,
\lk e^{-|p_0|r}\frac{T}{4\pi r}
\(\coth(\bar{r})-\frac1{\bar{r}}-(1-\delta_{p_0})\frac{\bar{r}}3\)\rk^2
+\order{\e} \;.
\end{align}
On the other hand, plugging \eqs\nr{eq:A6} and \nr{eq:A5} into the 
lhs of \eq\nr{eq:essence} results in
\begin{align}
\la{eq:lhs}
&\frac1{d\!-\!2}\sum_{p_0} 
\frac{T^3}{(4\pi)^2}
\int_0^\infty\!\!\!\!\!\!{\rm d}r\!\!
\int_0^\infty\!\!\!\!\!\!{\rm d}s\,
e^{-|p_0|(r+s)}
\(\coth\(\bar{r}\)-\frac1{\bar{r}}-(1\!-\!\delta_{p_0})\frac{\bar{r}}{3}\)
\(\coth\(\bar{s}\)-\frac1{\bar{s}}-(1\!-\!\delta_{p_0})\frac{\bar{s}}{3}\)
\times\nn&\times
\int\frac{{\rm d}^3\vec{p}}{(2\pi)^3}
\frac1{\vec{p}^2}\,\partial_r^2\,\frac{\sin(pr)}{pr}
\lk\frac{d}{\vec{p}^2}\,\partial_s^2+2\rk\frac{\sin(ps)}{ps}
+\order{\e}\;.
\end{align}
Using \eqs(H.21),(H.22) of \cite{Arnold:1994ps}, 
the integral in the last line of \eq\nr{eq:lhs} is
\begin{align}
\la{eq:deltas}
\frac1{2\pi}\(-\frac{d}{3r_>^3}+2\frac{\theta(r-s)}{r^3}\)
+\(d-2\)\frac{\delta(r-s)}{4\pi r s} 
&= \frac1{2\pi}\,\frac{3-d}{d}\,\frac{1}{3r_>^3}
+\(d-2\)\frac{\delta(r-s)}{4\pi r s} 
\end{align}
where $r_>=r\,\theta(r\!-\!s)\!+\!s\,\theta(s\!-\!r)$
and in the last step we used that the first line of \eq\nr{eq:lhs}
is symmetric under $r\leftrightarrow s$. 
Up to terms of $\order{\e}$, the second line of \eq\nr{eq:lhs} thus 
reduces\footnote{This is the ``amazing cancellation'' the authors of 
Ref.~\cite{Arnold:1994ps} refer to in their 
Appendix H.} to a delta function, transforming \eq\nr{eq:lhs} into 
\eq\nr{eq:target}, which completes the proof.

%
\end{appendix}

%

%
\end{document}